\documentclass[%
 reprint,
superscriptaddress,
 amsmath,amssymb,
 aps,
floatfix,
]{revtex4-2}
\usepackage{footnote}
\usepackage{graphicx}
\usepackage{epstopdf}
\usepackage{bm}
\usepackage{mathtools}
\usepackage{textcomp}
\usepackage{amsmath}
\usepackage{physics}
\usepackage{float}
\usepackage{siunitx}
\usepackage{amssymb}
\usepackage{float}
\usepackage{steinmetz}
\usepackage{csquotes}
\usepackage{lipsum}
\usepackage{soul,color}
\usepackage{braket}
\usepackage{accents}
\usepackage{url}
\usepackage{natbib}
\begin{document}

\preprint{APS/123-QED}
\title{Quantum emitter in a plasmonic field: an orientation generalised model}

\author{Harini Hapuarachchi}
\email[]{harini.hapuarachchi@rmit.edu.au}

\author{Jesse A. Vaitkus}

\author{Jared H. Cole}
\affiliation{School of Science, RMIT University, Melbourne, 3001, Australia}

\date{\today}

\begin{abstract}
When a quantum emitter (QE) is placed in close proximity to a plasmonic metal nanoparticle (MNP) within an external optical field, a dipole-dipole coupling arises, resulting in a highly tunable hybrid nanosystem that surpasses the optical manipulation capabilities of the individual components. These hybrid systems enable the exploration and manipulation of optical fields at the intersection between classical and quantum phenomena. Theoretical models of this interaction have typically been limited to the extreme orientations, where the QE and plasmonic dipoles are polarised either along or perpendicular to the inter-particle axis, for analytical tractability. In this work, we generalise the semi-classical optical dipole-dipole interaction model for a two-level quantum emitter in a plasmonic field for arbitrary polarisation angles. We show that the total field experienced by the quantum emitter at intermediate angles does not necessarily align with the external input field and discuss the implications of varying the polarisation angle of the external input field.

\end{abstract}

\maketitle
\section{Introduction} \label{Sec:Introduction}

Quantum emitters (QEs) are photon sources with properties that cannot be fully described by classical theories \cite{fox2023solid}. These include nanoscopic particles such as quantum dots \cite{harrison2016quantum, bera2010quantum}, different molecular and ionic systems \cite{fox2023solid} and colour centres in solids \cite{doherty2013nitrogen, aharonovich2022quantum}, that are crucial for a variety of applications in quantum information processing \cite{liu20192d, prawer2014quantum}, communication \cite{ren2022photonic, murtaza2023efficient}, optoelectronics \cite{yu2020quantum, litvin2017colloidal}, and sensing \cite{wu2016diamond, sapsford2006biosensing}. Light-matter interactions of many such quantum-confined systems (or their subsystems) can often be approximated using an effective two-level abstraction. That is, by considering the two electronic levels whose difference in energy is close to the interacting photon energy of the driving field \cite{novotny2012principles}.\vspace{0.75em}

Plasmonic metal nanoparticles (MNPs), such as gold and silver structures that are significantly smaller than the wavelength of incident light, exhibit cavity-like optical manipulation capabilities beyond the diffraction limit. This ability stems from the strong co-oscillation modes of their sea of conduction electrons coupled to the incoming coherent optical field. These modes, the dominant physics behind which is classically explainable, are known as localised surface plasmon resonances \cite{maier2007plasmonics, stockman2011nanoplasmonics}. \vspace{0.75em}

When a QE is placed at nanoscale proximity to an MNP, a dipole-dipole coupling arises between them, resulting in a highly tunable hybrid nanosystem that can significantly surpass the optical manipulation capabilities of the individual components \cite{govorov2006exciton, hartsfield2015single}. Therefore, QE-plasmonic hybrid systems, which allow the study and manipulation of physics at the interface of classical and quantum mechanics, have gained increasing research attention in the past few decades. A range of experiments have already demonstrated the ability to precisely synthesize and control such hybrid structures \cite{mertens2006polarization, lee2004bioconjugates, lee2005nanoparticle, hartsfield2015single, nisar2022controlling, reineck2013distance}.\vspace{0.75em}

The first semi-classical optical dipole-dipole interaction model for quantum emitter-plasmonic hybrid structures was published by A.O. Govorov \emph{et al.} in 2006 \cite{govorov2006exciton}. To date, the formalism they proposed has been used and adapted by an array of studies to interpret experiments involving QE-plasmonic hybrid systems \cite{cox2013nonlinear, persaud2020effect, nisar2022controlling}. Its adaptations have also been used to explore a range of phenomena including nonlinear-Fano effects \cite{artuso2008optical, kosionis2022coherent}, exciton-induced transparency \cite{artuso2010strongly, hatef2012plasmonic}, optical bistability \cite{malyshev2011optical, artuso2008optical}, gain without inversion \cite{sadeghi2010gain}, plasmonic metaresonances \cite{sadeghi2009plasmonic, hapuarachchi2019plasmonic}, and more \cite{kosionis2012nonlocal, yang2015ultrafast, lu2008enhancing, anton2013optical, mohammadzadeh2019resonance}. Such studies only consider extreme QE-plasmonic polarisations, where the external field (and therefore the QE and MNP dipoles) are polarised along, or perpendicular to, the MNP-QE axis.

Here we generalise the semi-classical optical dipole-dipole model for the interaction of a two-level quantum emitter with a plasmonic nanoparticle, capturing arbitrary polarisation angles.   

This paper is organised as follows: In section \ref{Sec:Formalism}, We derive the total field experienced by a quantum emitter in a MNP-QE system polarised at an arbitrary angle. For this, we assume that the complex exponential coefficient of the QE dipole aligns along the unit vector of the complex exponential coefficient of the total field experienced by the QE. We then express the time evolution of the QE state in a transformed reference frame (where the high-frequency dependence is eliminated), within the Lindblad formalism. In section \ref{Sec:Results_and_Discussion}, we convert the above equation of motion of the QE state (complex matrix differential equation) to a set of numerically solvable real-valued differential-algebraic equations (DAEs). We then verify the orientation-generalised model by comparing its output against the well-established extreme case solutions in the literature. We finally show that the total field experienced by the quantum emitter does not necessarily align along the external driving field for intermediate orientations and analyse the incident field polarisation dependence of QE populations.
	
\section{Formalism} \label{Sec:Formalism}

\subsection{System overview} \label{Sec:System_Overview}
\begin{figure}[t!]
	\includegraphics[width=0.65\columnwidth]{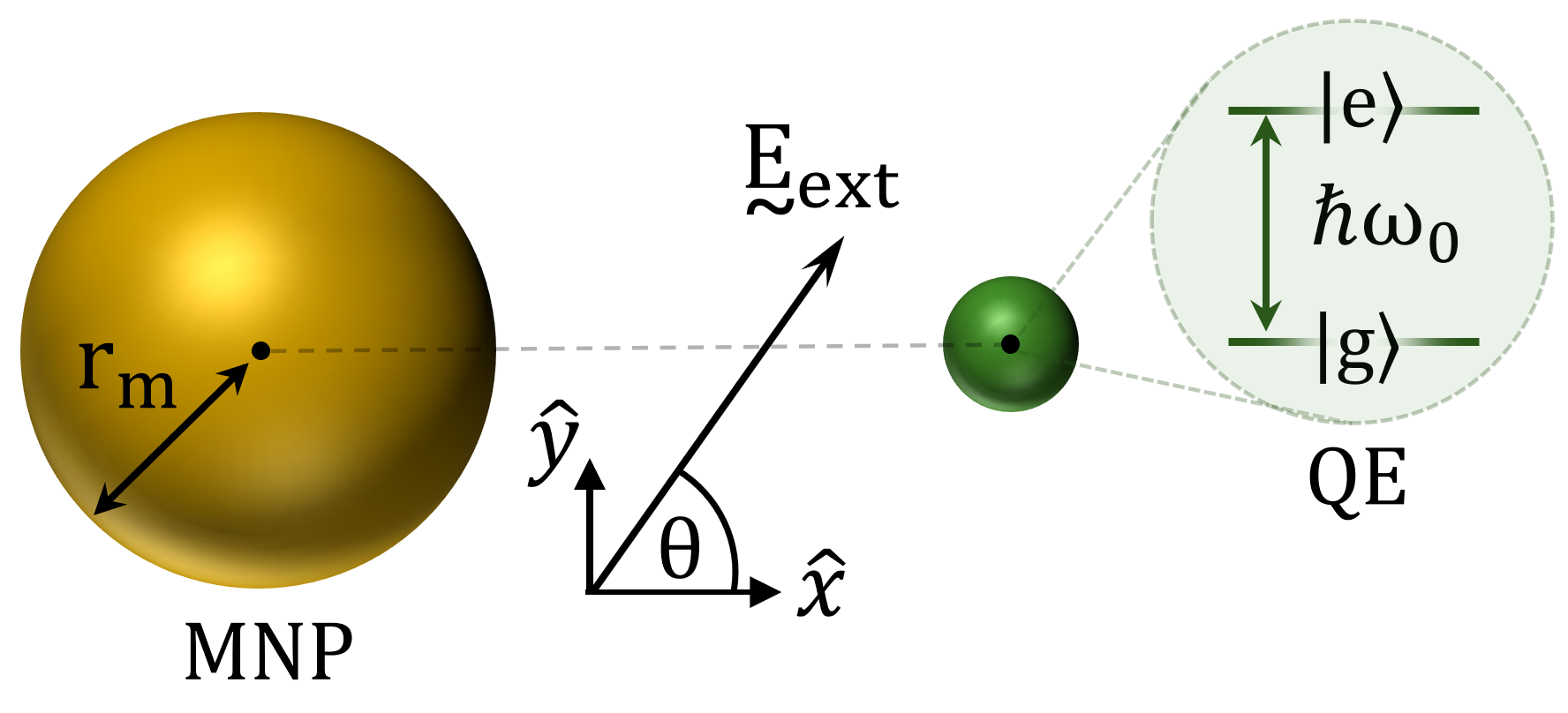}
	\centering
	\caption{(Colour online) Schematic illustration of the system under study. A two-level quantum emitter (QE), embedded in a material with relative permittivity $\epsilon_\text{q}$, undergoes dipole-dipole interaction with a plasmonic metal nanoparticle (MNP) of radius $r_m$ located at a centre separation $R$. An externally incident optical field electrically polarised at an angle $\theta$ with the MNP-QE hybrid axis is driving the system.\label{Fig:Schematic}}
\end{figure}

We consider a two-level quantum emitter (QE) with basis states $\lbrace\ket{g},\ket{e}\rbrace$ and exciton energy $\hbar\omega_0$ located at a centre separation $R$ from a spherical plasmonic metal nanoparticle (MNP) of radius $r_\text{m}$, as illustrated in Fig.\ \ref{Fig:Schematic}. Relative permittivities of the emitter material, plasmonic particle (resonator), and the background medium are denoted by $\epsilon_\text{q}, \epsilon_\text{m}(\omega)$, and $\epsilon_\text{b}$, respectively. The absolute permittivity of free-space is denoted by $\epsilon_0$. An external optical field with an electric component of the form, 
\begin{equation}\label{Eq:E_ext}
	\undertilde{E}_\text{ext} = \undertilde{E}_0 (e^{-i\omega t} + e^{i\omega t}),
\end{equation}
forming an angle $\theta$ with the MNP-QE axis is driving the composite system. In the above, $\omega$ and $t$ denote the angular frequency and time, respectively. The tilde ($\sim$) underneath denotes that the respective entity is a vector. 

The real-valued vector coefficient $\undertilde{E}_0$ can be decomposed into two components in the plane containing the MNP-QE axis and $\undertilde{E}_\text{ext}$ as follows,
\begin{equation}\label{eq:E0}
	\undertilde{E}_0 = E_0 \cos{\theta} \hat{x} + E_0 \sin{\theta} \hat{y},
\end{equation}
where $\hat{x}$ is the unit vector along the MNP-QE axis.

At the quasi-static limit where the composite system is significantly smaller than the wavelength of incident light, the phase of the harmonically oscillating field appears approximately spatially constant throughout the system, at a given time. In such cases, we can assume the simplified problem of the composite system in a spatially static field with a harmonic time dependence. This lowest order approximation suffices to describe nanoscopic systems of dimensions below $\sim \SI{100}{\nano\meter}$ for many purposes \cite{maier2007plasmonics}. 

Throughout this work, we adopt a convention where the spatially quasi-static coefficient of $e^{-i\omega t}$ of a time-harmonic entity $X$ is denoted by $\tilde{X}^+$. For the external field, we set $\undertilde{\tilde{E}}_\text{ext}^+ = \underaccent{\tilde}{E}_0$.

\subsection{Direct plasmonic response field} \label{Sec:Direct_plasmon_field}

Upon the incidence of $\undertilde{E}_\text{ext}$, an oscillatory dipole moment with an $e^{-i\omega t}$ coefficient of the following form is created in the plasmonic particle \cite{maier2007plasmonics},
\begin{align}\label{Eq:Direct_plasmonic_dipole}
	\underaccent{\tilde}{\tilde{d}}^+_\text{m} &= (4\pi\epsilon_0\epsilon_\text{b})\alpha(\omega)\underaccent{\tilde}{E}_0\\
	\text{where, }\alpha(\omega) &= r_\text{m}^3
	\left[\epsilon_\text{m}(\omega)-\epsilon_\text{b}\right]\big/\left[\epsilon_\text{m}(\omega)+2\epsilon_\text{b}\right].\nonumber
\end{align}

Using the point dipole approximation-based simplification procedure outlied in appendix \ref{Sec:Appendix_A} on the $\hat{x}$ and $\hat{y}$ components of (\ref{Eq:Direct_plasmonic_dipole}) separately, $e^{-i\omega t}$ coefficient of the direct plasmonic dipole response field experienced by the QE can be obtained as follows,
\begin{equation}\label{Eq:Direct_plasmon_res_field}
	\underaccent{\tilde}{\tilde{E}}^+_\text{m} = \frac{\alpha(\omega)E_0}{R^3\epsilon_\text{effQ}}(2\cos{\theta}\hat{x} -\sin{\theta}\hat{y}).
\end{equation}
In the above, $\epsilon_\text{effQ} = (2\epsilon_\text{b}+\epsilon_{q})\big/(3\epsilon_\text{b})$ accounts for the optical screening induced by the emitter material \cite{artuso2012optical,maier2007plasmonics}.

As plasmonic phenomena are typically ultrafast (femtosecond range) \cite{stockman2011nanoplasmonics}, we assume that the QE \enquote*{instantaneously} experiences the direct plasmonic response field above, as the external field is switched on.

\subsection{Self-feedback of the quantum emitter} \label{Sec:QE_self_feedback}

The classical equivalent of the transition dipole moment of the QE can be obtained as $\underaccent{\tilde}{d}_\text{q}=\mathrm{Tr}[\rho_\text{s}\underaccent{\tilde}{\mu}_\text{op}]$, where, $\rho_\text{s}$ is the density matrix in the static reference frame and $\underaccent{\tilde}{\mu}_\text{op} = \underaccent{\tilde}{\mu}_\text{ge}\ket{g}\!\bra{e} + \underaccent{\tilde}{\mu}_\text{eg}\ket{e}\!\bra{g}$ is the QE transition dipole operator \cite{yariv1967quantum, artuso2012optical}. This expression expands to,
\begin{equation}\label{Eq:QE_dipole}
	\underaccent{\tilde}{d}_\text{q} = \underaccent{\tilde}{\tilde{d}}_\text{q}^+e^{-i\omega t} + c.c. = \underaccent{\tilde}{\mu}_\text{ge}\tilde{\rho}_\text{eg}e^{-i\omega t} + c.c.,
\end{equation}
where $c.c.$ denotes the complex conjugate of the preceding expression. The high-frequency time dependence of the coherences of $\rho_\text{s}$ are factorised as $\rho_\text{eg}=\rho_\text{ge}^*=\tilde{\rho}_\text{eg}e^{-i\omega t}$.

When the quantum emitter is initialised in the ground state, the total field experienced by it ($\underaccent{\tilde}{E}_\text{tot}$) is only contributed by the screened external field and the direct plasmonic response field. That is,
\begin{equation}\label{Eq:E_tot_0}
	\underaccent{\tilde}{\tilde{E}}^+_\text{tot}(t=0)=\underaccent{\tilde}{E}_0/\epsilon_\text{effQ}+\underaccent{\tilde}{\tilde{E}}^+_\text{m}.
\end{equation}

As coherences of the QE gradually emerge, we assume that the $e^{-i\omega t}$ coefficient of the QE dipole aligns along the $e^{-i\omega t}$ coefficient of the total external field as follows,
\begin{align*}
	\underaccent{\tilde}{{\tilde{d}}}^+_\text{q}(t) = \underaccent{\tilde}{\mu}_\text{ge}(t) \;\tilde{\rho}_\text{eg}(t),\\
	\text{where, }\underaccent{\tilde}{\mu}_\text{ge}(t) =\mu_\text{ge}\hat{e}(t).
\end{align*}
In the above, $\hat{e}(t)$ is the complex-valued unit vector obtained by normalising the $e^{-i\omega t}$ coefficient of the total external field as $\underaccent{\tilde}{\tilde{E}}^+_\text{tot}(t) = C_x(t)\hat{x}+C_y(t)\hat{y}$, where $C_x(t)$ and $C_y(t)$ are complex-valued field components. This allows us to write,
\begin{align}\label{Eq:d_q_t_greater_than_0}
	&\underaccent{\tilde}{{\tilde{d}}}^+_\text{q}(t) = \tilde{d}^{x+}_\text{q}(t) \hat{x} + \tilde{d}^{y+}_\text{q}(t) \hat{y}, \text{ where,}\\
	\tilde{d}^{j+}_\text{q}(t) &=\frac{\mu_{ge}\tilde{\rho}_\text{eg}(t) C_j(t)}{\sqrt{ |C_x(t)|^2+ |C_y(t)|^2}}, \text{ for } j\in\lbrace x,y\rbrace.\nonumber
\end{align}
Applying equation (\ref{Eq:Dipole_response_extreme}) on the $\hat{x}$ and $\hat{y}$ components of (\ref{Eq:d_q_t_greater_than_0}) separately, we obtain the quasi-static coefficient of the screened QE dipole response field experienced at the centre of MNP as,
\begin{equation}\label{Eq:QE_response_field}
	\underaccent{\tilde}{\tilde{E}}_\text{qm}^+ = \frac{2\tilde{d}^{x+}_\text{q}(t)\hat{x} -\tilde{d}^{y+}_\text{q}(t)\hat{y}}{4\pi\epsilon_0\epsilon_\text{b}R^3\epsilon_\text{effQ}}
\end{equation}

The response dipole induced in the MNP due to the above field can be obtained similarly to (\ref{Eq:Direct_plasmonic_dipole}). It has an $e^{-i\omega t}$ coefficient,
\begin{equation}\label{Eq:QE_feedback_dipole_in_MNP}
	\underaccent{\tilde}{\tilde{d}}^+_\text{qm} = 4\pi\epsilon_0\epsilon_\text{b}\alpha(\omega)\underaccent{\tilde}{\tilde{E}}_\text{qm}^+.
\end{equation}
This additional plasmonic dipole induces a \enquote*{self-feedback} field of the following form at the QE location,
\begin{multline}\label{Eq:QE_self_Feedback}
	\underaccent{\tilde}{\tilde{E}}_\text{qs}^+(t) = \tilde{E}_\text{qs}^{x+}(t) \; \hat{x} + \tilde{E}_\text{qs}^{y+}(t) \; \hat{y}\\
	=d_\text{Qcoeff} [4 \tilde{d}^{x+}_\text{q}(t) \hat{x} + \tilde{d}^{y+}_\text{q}(t)\hat{y}], 
\end{multline}
which has been obtained by applying (\ref{Eq:Dipole_response_extreme}) on (\ref{Eq:QE_feedback_dipole_in_MNP}), while accounting for screening. In the above, we have defined,
\begin{equation}
	d_\text{Qcoeff} = \frac{\alpha(\omega)}{4\pi\epsilon_0\epsilon_\text{b}R^6\epsilon_\text{effQ}^2}.
\end{equation}

\subsection{Total field experienced by the QE} \label{Sec:Total_field_on_QE}
The total field $\underaccent{\tilde}{E}_\text{tot}(t)$ contributing to exciton formation in the QE comprises the external field, direct plasmonic response field, and self-feedback through the plasmonic particle. Thus, using (\ref{Eq:E_ext}), (\ref{Eq:Direct_plasmon_res_field}), and (\ref{Eq:QE_self_Feedback}), we obtain,

\begin{equation}\label{Eq:E_tot}
	\underaccent{\tilde}{\tilde{E}}_\text{tot}^+(t) = C_x(t)\hat{x} + C_y(t)\hat{y},
\end{equation}
where,
\begin{subequations}\label{Eq:Cx_and_C_y}
	\begin{align}
	C_x(t) &= C_x(0) + \tilde{E}_\text{qs}^{x+}(t),\\
	C_y(t) &= C_y(0) + \tilde{E}_\text{qs}^{y+}(t), 
	\end{align}
\end{subequations}
with initial values of these components at $t$=0 given by,
\begin{subequations}\label{Eq:Cx0_and_C_y0}
	\begin{align}
		C_x(0) &= \frac{E_0\cos{\theta}}{\epsilon_\text{effQ}}\left(1+\frac{2\alpha(\omega)}{R^3}\right), \\
		C_y(0)&=\frac{E_0\sin{\theta}}{\epsilon_\text{effQ}}\left(1-\frac{\alpha(\omega)}{R^3}\right).
	\end{align}
\end{subequations}

\subsection{Hamiltonian of the quantum emitter} \label{Sec:Hamiltonian}\vspace{-0.5em}
At the large photon limit where the optical field behaves classically, the Hamiltonian of the QE in the static reference frame takes the following form \cite{artuso2010strongly, artuso2012optical, kosionis2012nonlocal, hapuarachchi2020influence, steinfeld2023prospects},
\begin{equation}\label{Eq:Unperturbed_Hamiltonian}
	H_\text{s} = \hbar\omega_0\sigma_+\sigma_- - \underaccent{\tilde}{\mu}_\text{op}(t)\cdot \underaccent{\tilde}{E}_\text{tot}(t).
\end{equation}
In the above, $\sigma_+=\ket{e}$$\bra{g}$ and $\sigma_-=\ket{g}$$\bra{e}$ are the raising and lowering operators of the QE. The zero energy level is defined at $\ket{g}$. 

For the ease of computations, we transform $H_\text{s}$ into an alternative reference frame where the high-frequency time-dependence of the QE state is eliminated, using the following expression \cite{messiah2014quantum} (derivation of which is outlined in Appendix \ref{Sec:Appendix_B}),

\begin{equation}\label{Eq:Unitary_transformation}
	H_\text{t} = U_\text{op} H_\text{s} U_\text{op}^\dagger + i\hbar\dot{U}_\text{op}U_\text{op}^\dagger.
\end{equation}
The unitary transformation operator $U_\text{op}$ takes the following form.
\begin{equation}\label{Eq:Unitary_op}
	U_\text{op} = e^{iH_\text{u}t/\hbar}, \text{where  }H_\text{u} = \hbar\omega\ket{e}\!\bra{e}.\nonumber
\end{equation}
As $H_\text{u}$ possesses the eigenstates $\ket{a}\in\lbrace\ket{g},\ket{e}\rbrace$ with eigenvalues $\hbar\omega_a\in\lbrace 0,\hbar\omega\rbrace$, we can write,
\begin{equation*}\label{Eq:Exponent_eqs}
	e^{\pm i H_\text{u}t/\hbar}\ket{a}=e^{\pm i \omega_a t}\ket{a}.
\end{equation*}
Using the above expressions and the rotating wave approximation to simplify (\ref{Eq:Unitary_transformation}) we arrive at,
\begin{equation}\label{Eq:H_t_final}
	H_\text{t} \approx \hbar(\omega_0-\omega)\sigma_+\sigma_- - \hbar(\Omega_\text{eff}\sigma_+ + \Omega_\text{eff}^*\sigma_-),
\end{equation}
where the effective Rabi frequency $\Omega_\text{eff}$ is defined as,
\begin{align}\label{Eq:Rabi_eff}
	&\;\;\;\;\;\;\;\;\;\;\;\;\Omega_\text{eff} = \Omega + \eta\tilde{\rho}_\text{eg}(t),\text{ where,}\\ 
	&\Omega = \frac{\mu_\text{ge}}{\hbar}\left(\frac{C_x(0) C_x^*(t)+ C_y(0) C_y^*(t)}{\sqrt{|C_x(t)|^2+|C_y(t)|^2}} \right),\nonumber\\
	\eta = &\frac{\mu_\text{ge}^2 d_\text{Qcoeff}}{\hbar}\left( \frac{4|C_x(t)|^2 + |C_y(t)|^2}{\sqrt{|C_x(t)|^2+|C_y(t)|^2}}  \right). \nonumber
\end{align}

\subsection{Density matrix in the transformed reference frame}\label{Sec:Density_matrix_in_RotFrame}
It is necessary to transform the density matrix (state) of the QE to the same reference frame as its Hamiltonian prior to solving for the dynamics of the system. The density matrix in the static reference frame can be expressed in the $\lbrace\ket{g}, \ket{e}\rbrace$ basis as,
\begin{equation*}
	\rho_\text{s} = \rho_\text{gg}\ket{g}\!\bra{g} + \tilde{\rho}_\text{ge} e^{i\omega t}\ket{g}\!\bra{e} + \tilde{\rho}_\text{eg} e^{-i\omega t}\ket{e}\!\bra{g} + \rho_\text{ee}\ket{e}\!\bra{e}.
\end{equation*}

The density matrix in the transformed reference frame $\rho_t$ can be expressed as a statistical mixture of the pure states $\ket{\psi_t}_j = U_\text{op}\ket{\psi_s}_j$ with classical probabilities $p_j$, where $\lbrace\ket{\psi_s}_j\rbrace$ denote pure states in the static reference frame. It can be simplified as follows:
\begin{equation}
	\rho_t = \sum_j p_j \ket{\psi_t}_j\!\bra{\psi_t}_j	= U_\text{op}\rho_\text{s}U_\text{op}^\dagger=\begin{pmatrix}
		\rho_\text{gg}&\tilde{\rho}_\text{ge}\\
		\tilde{\rho}_\text{eg}&\rho_\text{ee}
	\end{pmatrix}.
\end{equation}

\subsection{Open quantum dynamics}\label{Sec:Open_quantum_dynamics}
The Hamiltonian of the QE, coherently interacting with the total field incident on it, represents a closed quantum system where the incoherent impact of the environment is yet to be taken into account. We assume that the QE weakly couples to the environment resulting in an open quantum system with irreversible dynamics. Here we estimate the evolution of the QE as an open quantum system within the Lindblad formalism as follows \cite{breuer2002theory, campaioli2024quantum},
\begin{equation}\label{Eq:Master_equation}
	\dot{\rho}_t = \frac{i}{\hbar}\left[\rho_t,H_t\right] + \sum_{k=1}^2 \lambda_k \left[2 L_k \rho_t L_k^\dagger - \lbrace L_k^\dagger L_k, \rho_t \rbrace\right], 
\end{equation}
where we account for the incoherent relaxation (with $L_1=\ket{g}\!\bra{e}$ and $\lambda_1$) and dephasing (with $L_2=\ket{e}\!\bra{e}$ and $\lambda_2$) of the QE excitation. The mathematical operations $[\cdot,\cdot]$ and $\lbrace\cdot,\cdot\rbrace$ denote commutator and anti-commutator of the operands, respectively. 

By element-wise comparison of the left and right hand sides of (\ref{Eq:Master_equation}), we arrive at the following optical Bloch equations for the quantum emitter \cite{hapuarachchi2020influence, artuso2012optical},
\begin{subequations}\label{Eq:Optical_Bloch_eqs}
\begin{align}
	\dot{\rho}_\text{ee} &= -\frac{\rho_\text{ee}}{T_1} + i\Omega_\text{eff}\tilde{\rho}_\text{ge}-i\Omega_\text{eff}^*\tilde{\rho}_\text{eg},\label{Eq:Optical_Bloch_complex_eq1}\\
	\dot{\rho}_\text{gg} &= \frac{\rho_\text{ee}}{T_1} - i\Omega_\text{eff}\tilde{\rho}_\text{ge}+i\Omega_\text{eff}^*\tilde{\rho}_\text{eg},\label{Eq:Optical_Bloch_complex_eq2}\\
	\dot{\tilde{\rho}}_{eg} &= -[i(\omega_0-\omega)+1/T_2]\tilde{\rho}_\text{eg} + i\Omega_\text{eff}(\rho_\text{gg}-\rho_\text{ee}).\label{Eq:Optical_Bloch_complex_eq3}
\end{align}
\end{subequations}
where $T_1 = 1/(2\lambda_1)$ is the population decay time and $T_2=1/(\lambda_1+\lambda_2)$ is the coherence time of the QE \cite{fox2023solid}.

\section{Results and discussion}\label{Sec:Results_and_Discussion}

\subsection{Numerical solution procedure and parameters} \label{Sec:Numerical_procedure}

For the the purpose of efficiently solving numerically, the above complex-valued differential equations can be recast to a real-valued form as \cite{hapuarachchi2020influence, artuso2012optical}, 
\begin{subequations}\label{Eq:Real_diff_eqs}
	\begin{align}
		\dot{\mathcal{A}} &= -\frac{\mathcal{A}}{T_2} + \delta \mathcal{B} - (\Omega_\text{im} + \eta_\text{im}\mathcal{A} - \eta_\text{re}\mathcal{B})\Delta,\\
		\dot{\mathcal{B}} &= -\frac{\mathcal{B}}{T_2} - \delta \mathcal{A} - (\Omega_\text{re} + \eta_\text{re}\mathcal{A} + \eta_\text{im}\mathcal{B})\Delta,\\
		\dot{\Delta} &= \frac{1-\Delta}{T_1} + 4\left[\Omega_\text{im}\mathcal{A} + \Omega_\text{re}\mathcal{B}+\eta_\text{im}(\mathcal{A}^2 + \mathcal{B}^2)\right],
	\end{align}
\end{subequations}
where $\Delta = \rho_\text{gg}-\rho_\text{ee}$ denotes the population difference, $\delta = \omega-\omega_0$ denotes the detuning, $\tilde{\rho}_\text{ge}=\mathcal{A}+i\mathcal{B}$, $\Omega=\Omega_\text{re}+i\Omega_\text{im}$, and $\eta = \eta_\text{re}+i\eta_\text{im}$.

From the above definitions and (\ref{Eq:Rabi_eff}), it becomes evident that the coefficients of (\ref{Eq:Real_diff_eqs}) are reliant on $C_x(t)$ and $C_y(t)$. This necessitates solving the above differential equations alongside (\ref{Eq:Cx_and_C_y}). For this we define,
\begin{subequations}
	\begin{align*}
		&C_j(t) = C_\mathrm{j,re} + iC_\mathrm{j,im},\\
		&C_j(0) = C_\mathrm{j,re}^0 + iC_\mathrm{j,im}^0, \\
		&d_\text{Qcoeff} = d_\text{re} + i d_\text{im},
	\end{align*}
\end{subequations}
for $j\in\lbrace x,y\rbrace$. We then decompose (\ref{Eq:Cx_and_C_y}) as,
\begin{widetext}
	\begin{subequations}\label{Eq:Real_algebraic_eqs}
		\begin{align}
			C_\mathrm{x,re} &= C_\mathrm{x,re}^0 + 4(\mu_\text{ge}\big/M)\left[(d_\text{re}C_\mathrm{x,re} - d_\text{im}C_\mathrm{x,im})\mathcal{A} +(d_\text{re}C_\mathrm{x,im} + d_\text{im}C_\mathrm{x,re})\mathcal{B} \right],\\
			C_\mathrm{x,im} &= C_\mathrm{x,im}^0 + 4(\mu_\text{ge}\big/M)\left[(d_\text{re}C_\mathrm{x,im} + d_\text{im}C_\mathrm{x,re})\mathcal{A} -(d_\text{re}C_\mathrm{x,re} - d_\text{im}C_\mathrm{x,im})\mathcal{B} \right],\\
			C_\mathrm{y,re} &= C_\mathrm{y,re}^0 + (\mu_\text{ge}\big/M)\left[(d_\text{re}C_\mathrm{y,re} - d_\text{im}C_\mathrm{y,im})\mathcal{A} +(d_\text{re}C_\mathrm{y,im} + d_\text{im}C_\mathrm{y,re})\mathcal{B} \right],\\
			C_\mathrm{y,im} &= C_\mathrm{y,im}^0 + (\mu_\text{ge}\big/M)\left[(d_\text{re}C_\mathrm{y,im} + d_\text{im}C_\mathrm{y,re})\mathcal{A} -(d_\text{re}C_\mathrm{y,re} - d_\text{im}C_\mathrm{y,im})\mathcal{B} \right],\\
			M & = \sqrt{C_\mathrm{x,re}^2+C_\mathrm{x,im}^2+C_\mathrm{y,re}^2+C_\mathrm{y,im}^2},
		\end{align}
	\end{subequations}
\end{widetext}
Together, (\ref{Eq:Real_diff_eqs}) and (\ref{Eq:Real_algebraic_eqs}) form a set of real-valued differential algebraic equations (DAEs). 

To obtain the numerical results presented in this work, we solved this DAE system using a class-based approach combining a \emph{Runge-Kutta} solver with the \emph{fsolve()} root finding method in Python. The full implementation of our code is provided in the supplementary material \cite{SuppmatCode}. We verified that this approach works well for relatively low and moderate nonlinearity regimes at reasonably small detunings, such as the parameter space explored in this work. Alternative numerical solution approaches may need to be sought for high nonlinearity regimes. A detailed discussion of nonlinearity regimes of exciton-plasmon hybrids can be found in \cite{artuso2008optical}. 

Common parameters used in this study are as follows: We consider a QE interacting with a gold nanoparticle kept at a centre separation of $R=\SI{25}{\nano\meter}$ in air ($\epsilon_\text{b} = 1$). Relative permittivity of the gold nanoparticle $\epsilon_\text{m}(\omega)$ is obtained via cubic-spline interpolation of the tabulations by John and Christy \cite{johnson1972optical}. Energy of the QE ($\hbar\omega_0$) is set to be equal to the plasmonic resonance $\approx\SI{2.355}{\electronvolt}$. Relative permittivity of the QE host material, exciton relaxation time, and ensemble dephasing time are set to $\epsilon_\text{q}=6$, $T_1=\SI{0.8}{\nano\second}$, and $T_2=\SI{0.3}{\nano\second}$, respectively \cite{artuso2008optical, govorov2006exciton, kosionis2012nonlocal}. We consider a real transition dipole element $\mu_\text{ge}= 0.2$ e.nm \cite{kosionis2012nonlocal}. The $e^{-i\omega t}$ coefficient of the incoming field $E_0=\SI{3e4}{\volt\per\meter}$, resulting in an intensity of $I_0 = 2\sqrt{\epsilon_\text{b}}\epsilon_0 c E_0^2 \sim \SI{5e2}{\watt\per\square\centi\meter}$ in air \cite{guest2002measurement, kosionis2012nonlocal}. 

For analytical tractability, we use a constant value for the relaxation time, in line with the practice widely adopted in the literature \cite{govorov2006exciton, sadeghi2009plasmonic, hatef2012coherent, hatef2012quantum, artuso2008optical, artuso2010strongly, hapuarachchi2018exciton, hapuarachchi2020influence, kosionis2012nonlocal}. Potential future extensions of this work includes accounting for the plasmonic influence on the exciton relaxation rate. This requires generalising the derivations for plasmon induced emitter decay rate modifications at extreme polarisations (such as those by \emph{Carminati et al.} \cite{carminati2006radiative}) for arbitrary polarisation angles, which is beyond the scope of this work.

\subsection{Verification of the formalism} \label{Sec:Verification}

\begin{figure}[t!]
	\includegraphics[width=\columnwidth]{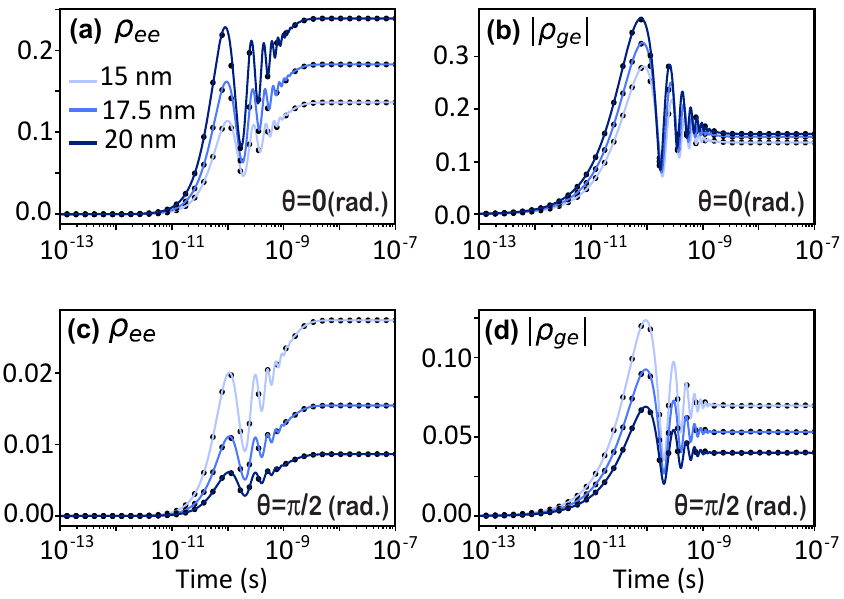}
	\centering
	\caption{(Colour online) Verification of the angle generalised solution procedure against the extreme-case solutions in the literature for plasmonic radii $r_\text{m}\in\lbrace15, 17.5,20\rbrace\SI{}{\nano\meter}$ and external field detuning $\delta=\SI{20}{\micro\electronvolt}$. Solid lines in blue shades denote the excited state populations ($\rho_\text{ee}$) and coherence magnitudes ($|\rho_\text{ge}|$) obtained using the complex-unit vector based angle-generalised solution procedure proposed in this work. Black dots denote the respective solutions obtained using the extreme-case solution procedure established in the literature, outlined in section \ref{Sec:Verification}. \label{Fig:Verification}}
\end{figure}

The approach widely adopted in the literature to solve for the extreme cases of the semi-classical dipole-dipole interaction of QE-plasmonic hybrids is to define an orientation parameter $s_\alpha$, with values $2$ and $-1$ for the axial ($\theta=\SI{0}{\radian}$) and tangential ($\theta=\pi/\SI{2}{\radian}$) cases, respectively. The effective Rabi frequency components are then defined as follows \cite{sadeghi2009plasmonic, hatef2012coherent, hatef2012quantum, artuso2008optical, artuso2010strongly, hapuarachchi2018exciton, hapuarachchi2020influence, kosionis2012nonlocal},
\begin{subequations}\label{Eq:Rabi_eff_extreme_components}
	\begin{align}
		\Omega\big|_{\theta\in\lbrace0,\pi/2\rbrace} &= \frac{\mu_\text{ge} E_0}{\hbar\epsilon_\text{effQ}}\left(1 + \frac{s_\alpha\alpha(\omega)}{R^3}\right),\\
		\eta\big|_{\theta\in\lbrace0,\pi/2\rbrace} &= \frac{s_\alpha^2 \mu_\text{ge}^2 \alpha(\omega)}{4\pi\epsilon_0\epsilon_\text{b}\hbar\epsilon_\text{effQ}^2 R^6},
	\end{align}
\end{subequations}
when solving QE optical Bloch equations in (\ref{Eq:Optical_Bloch_eqs}).

In Fig.\ \ref{Fig:Verification}, we verify the angle-generalised model derived in this work by solving the set of DAEs defined in  (\ref{Eq:Real_diff_eqs}) and (\ref{Eq:Real_algebraic_eqs}), where the the effective Rabi frequency is given by (\ref{Eq:Rabi_eff}), against the above extreme case solutions established in the literature (\ref{Eq:Rabi_eff_extreme_components}). It is evident that the generalised model successfully reproduces the extreme case solutions.

\subsection{Dynamics of the total field incident on the QE} \label{Sec:Dynamics_of_total_field}

\begin{figure}[t!]
	\includegraphics[width=0.85\columnwidth]{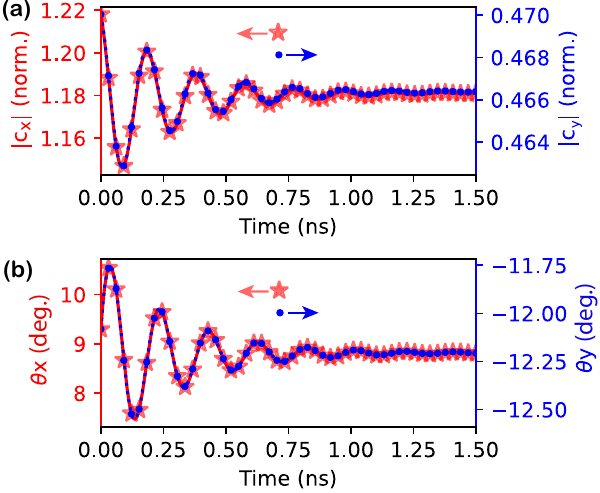}
	\centering
	\caption{(Colour online) Time-domain variation of the total optical field experienced by the quantum emitter placed next to a spherical plasmonic nanoparticle of radius $\SI{15}{\nano\meter}$, upon the incidence of an external field at an angle $\theta=\pi/4\;\SI{}{\radian}$ and a detuning of $\delta=\SI{20}{\micro\electronvolt}$. (a) Magnitudes of the $\hat{x}$ and $\hat{y}$ directional components of the $e^{-i\omega t}$ coefficient of the total optical field experienced by the QE in the presence of the plasmonic nanoparticle, normalised by the field amplitude experienced in the absence of the latter. (b) Plasmonically induced phase lags in the $\hat{x}$ and $\hat{y}$ directional components of the total field incident on the quantum emitter. \label{Fig:Field_for_45_deg}}
\end{figure}

After verification of the angle generalised solution procedure, we examine the time evolution of the plasmonically modified total field incident on the quantum emitter  for an intermediate orientation of external field incidence. For this, we express the total field incident on the QE as,
\begin{multline}\label{Eq:Total_field_full_form}
	\underaccent{\tilde}{E}_\text{tot} = 2|C_x(t)|\cos\left[\omega t-\theta_x(t)\right]\hat{x} \\
	+ 2|C_y(t)|\cos\left[\omega t-\theta_y(t)\right]\hat{y},
\end{multline}
utilising the following polar forms of the $\hat{x}$ and $\hat{y}$ components of $\underaccent{\tilde}{\tilde{E}}_\text{tot}^+$ ($e^{-i\omega t}$ component of $\underaccent{\tilde}{E}_\text{tot}$) in (\ref{Eq:E_tot}),
\begin{subequations}
	\begin{align}
		C_x(t) &= |C_x(t)|e^{i\theta_x(t)},\\
		C_y(t) &= |C_y(t)|e^{i\theta_y(t)}.
	\end{align}
\end{subequations}
In the absence of the plasmonic particle, the external field (shielded by the emitter material) experienced by the isotropic QE takes the following form,
\begin{align}\label{Eq:E_iso}
	\underaccent{\tilde}{E}_\text{iso} &= \frac{\underaccent{\tilde}{E}_0}{\epsilon_\text{effQ}}(e^{-i\omega t}+e^{i\omega t})\nonumber\\
	&=\frac{2E_0\cos(\omega t)}{\epsilon_\text{effQ}}\left(\cos\theta \hat{x} + \sin\theta \hat{y} \right).
\end{align}
We plot the $\hat{x}$ and $\hat{y}$ amplitudes of $\underaccent{\tilde}{\tilde{E}}_\text{tot}^+$ (normalised by the $e^{-i\omega t}$ coefficient $\underaccent{\tilde}{E}_\text{iso}$) and plasmonically-induced phase lags $\theta_x(t)$ and $\theta_y(t)$, for a case where the incident field forms an angle $\theta=\pi/4\;\text{rad}$ with the QE-MNP axis, in Fig.\ \ref{Fig:Field_for_45_deg}.

It can be observed that the $\hat{x}$ and $\hat{y}$ directional amplitudes and phase-lags of $\underaccent{\tilde}{E}_\text{tot}$ initiate at finite values attributable to $C_x(0)$ and $C_y(0)$ in (\ref{Eq:Cx0_and_C_y0}) and undergo an oscillatory time variation prior to settling at their respective steady state values. Notice that in the example considered, the steady state amplitude values along both $\hat{x}$ and $\hat{y}$ are lesser than the respective initial amplitudes. This may, at a glance, seem counter-intuitive as the QE self-feedback components $\tilde{E}_\text{qs}^{x+}(t)$ and $\tilde{E}_\text{qs}^{y+}(t)$ gradually \enquote*{add} to the initial values of $C_x(t)$ and $C_y(t)$ with time, as per (\ref{Eq:Cx_and_C_y}). It is noteworthy that this \enquote*{addition} of $e^{-i\omega t}$ coefficients corresponds to a superposition of those (classical) optical field components, which could be either constructive or destructive in different regions.

In our example where $\theta=\pi/4\;\text{rad}$, the isolated quantum emitter experiences equally-valued $\hat{x}$ and $\hat{y}$ directional amplitudes ($\sqrt{2}E_0\big/\epsilon_\text{effQ}$), as evident from (\ref{Eq:E_iso}). However, it can be observed from Fig.\ \ref{Fig:Field_for_45_deg} and (\ref{Eq:Total_field_full_form}) that the $\hat{x}$ and $\hat{y}$ directional amplitudes of $\underaccent{\tilde}{E}_\text{tot}$ experienced by the QE in the presence of a plasmonic particle are differently valued, irrespective of their patterns of time-variation being similar. Therefore, the plasmonically influenced total field experienced by the QE does not necessarily align along the external input field ($\underaccent{\tilde}{E}_\text{ext} = \underaccent{\tilde}{E}_0(e^{-i\omega t}+e^{i\omega t})$) for intermediate orientations. 

\subsection{Impact of varied excitation angle}\label{Sec:Impact_of_varied_theta}

\begin{figure}[t!]
	\includegraphics[width=0.9\columnwidth]{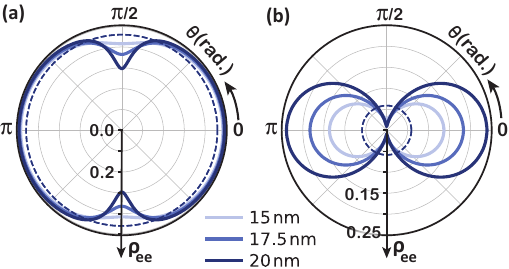}
	\centering
	\caption{(Colour online) Polar plots of the quantum emitter's excited state population ($\rho_\text{ee}$) at the steady state, in the presence of plasmonic gold nanoparticles with different radii, $r_\text{m}\in\lbrace\SI{15}{\nano\meter},\SI{17.5}{\nano\meter},\SI{20}{\nano\meter}\rbrace$ (a) when the external field detuning  $\delta=0$, and (b) when $\delta=\SI{20}{\micro\electronvolt}$. The dashed curves represent the excited population $\rho_\text{ee}$ of the isolated isotropic quantum emitter, in each case. \label{Fig:Polarrm}}
\end{figure}

In Fig.\ \ref{Fig:Polarrm}, we examine the behaviour of the steady state excited population $\rho_\text{ee}$ of the quantum emitter, which is proportional to both optical absorption \cite{artuso2008optical} and emission \cite{carmichael2013statistical}. Excited state populations in the presence of plasmonic nanoparticles with different radii are compared against that of the isolated quantum emitter.
An anisotropy resulting from the plasmon-assisted total field enhancement along the MNP-QE axis ($\hat{x}$) and suppression along the perpendicular axis ($\hat{y}$) is clearly visible. 

In Fig.\ \ref{Fig:Polarrm}(a), where the external field is resonant with the QE exciton energy, the excited state population is nearly saturated. Only a minor enhancement of $\rho_\text{ee}$ can be seen for a large fraction of orientations $\theta$ around the axial ($\theta=0, \pi$ $\SI{}{\radian}$) cases, for all plasmonic radii considered.
For the perpendicular orientations (where $\theta=\pm\pi/2$ $\SI{}{\radian}$), larger plasmonic particles result in higher levels of optical field suppression leading to larger $\rho_\text{ee}$ anisotropy. In Fig.\ \ref{Fig:Polarrm}(b), the external field is detuned from the QE exciton energy ($\delta=\SI{20}{\micro\electronvolt}$) and $\rho_\text{ee}$ is far from saturation. In this case, significant plasmon-assisted $\rho_\text{ee}$ enhancement that increases with the plasmonic nanoparticle size is observed along the axial dimension. Such regions of significant anisotropy hold promise for input field polarisation-based switching applications.

\section{Summary and Conclusion}
We generalised the semi-classical dipole-dipole interaction model for a quantum emitter (QE) coupled to a plasmonic metal nanoparticle (MNP) for all polarisations of the incoming optical field. Our mathematical abstraction relies on aligning the complex exponential coefficient of the quantum emitter's transition dipole along the complex exponential coefficient of the plasmonically modified total field experienced by the QE. This approach was verified by comparing its solutions for extreme optical polarisations against the orientation parameter-based extreme-case solution procedure established in the literature. We showed that the total field experienced by the quantum emitter does not necessarily align along the external input field, for intermediate polarisations. Our simulations also reveal higher levels of anisotropy of QE populations (and therefore of absorption and emission) for detuned driving fields where the QE is yet to be saturated. This holds promise for input field polarisation-based switching applications. 

\section*{Acknowledgements}
HH gratefully acknowledges RMIT University's Vice-Chancellor's Postdoctoral Research Fellowship and Australian Research Council's Centre of Excellence in Exciton Science (grant number CE170100026) for funding,  National Computational Infrastructure (NCI) supported by the Australian Government for computational resources, Francesco Campaioli for useful discussions on the numerical solution procedure and Dinuka U. Kudavithana for encouragement and support.

\renewcommand{\theequation}{A-\arabic{equation}}
  \setcounter{equation}{0}  

\appendix
\section{Plasmon response field components}\label{Sec:Appendix_A}
The dipole response field $\underaccent{\tilde}{E}(\underaccent{\tilde}{r})$ of a point dipole $\underaccent{\tilde}{d}$ on the origin experienced at a position $\underaccent{\tilde}{r}=|\underaccent{\tilde}{r}|\hat{r}$ can be obtained as \cite{griffiths1999introduction},
\begin{equation}
	\underaccent{\tilde}{E}(\underaccent{\tilde}{r}) = \frac{1}{4\pi\epsilon_0\epsilon_\text{b}|\underaccent{\tilde}{r}|^3}\left[3(\underaccent{\tilde}{d}\cdot \hat{r})\hat{r}-\underaccent{\tilde}{d}\right].
\end{equation}
For cases where $\underaccent{\tilde}{d}$ is parallel (perpendicular) to $\hat{r}$, the above expression simplifies to \cite{hapuarachchi2020influence},
\begin{equation}\label{Eq:Dipole_response_extreme}
	\underaccent{\tilde}{E}(\underaccent{\tilde}{r}) = \frac{s_\alpha \underaccent{\tilde}{d}}{4\pi\epsilon_0\epsilon_\text{b}|\underaccent{\tilde}{r}|^3}
\end{equation}
where the orientation parameter $s_\alpha=2$ (-1) for parallel (perpendicular) polarisation.

\renewcommand{\theequation}{B-\arabic{equation}}
\setcounter{equation}{0}  

\section{Unitary transformation}\label{Sec:Appendix_B}
Let unitary operator $U_\text{op}$ transform a pure state $\ket{\psi_s}$ to another pure state $\ket{\psi_t}$ in an alternate reference frame as follows,
\begin{equation}\label{Eq:unitary_kets}
	\ket{\psi_t} = U_\text{op}\ket{\psi_s} \implies \ket{\psi_s} = U^\dagger_\text{op}\ket{\psi_t}
\end{equation}
Evolution of both $\ket{\psi_s}$ and $\ket{\psi_t}$ are governed by the Schr$\ddot{\mathrm{o}}$dinger's equation, with the Hamiltonian in the respective reference frame,
\begin{equation}\label{Eq:Schrodinger_static}
i\hbar\frac{\partial}{\partial t}\ket{\psi_s} = H_s\ket{\psi_s}
\end{equation}
\begin{equation}\label{Eq:Schrodinger_transformed}
	i\hbar\frac{\partial}{\partial t}\ket{\psi_t} = H_t\ket{\psi_t}.
\end{equation}
Substituting for $\ket{\psi_t}$ in (\ref{Eq:Schrodinger_transformed}) from (\ref{Eq:unitary_kets}), 
\begin{align*}
	i\hbar\frac{\partial}{\partial t}(U_\text{op}\ket{\psi_s}) &= H_t(U_\text{op}\ket{\psi_s})\\
	\implies i\hbar(\frac{\partial}{\partial t}U_\text{op}\ket{\psi_s} &+ U_\text{op}\frac{\partial}{\partial t}\ket{\psi_s}) = H_t U_\text{op}\ket{\psi_s}
\end{align*}
Substituting from (\ref{Eq:Schrodinger_static}), extracting the coefficients of $\ket{\psi_s}$, and right multiplying by $U_\text{op}^\dagger$ throughout, the following formula for the transformed Hamiltonian can be obtained,
\begin{equation}
	H_\text{t} = U_\text{op} H_\text{s} U_\text{op}^\dagger + i\hbar\dot{U}_\text{op}U_\text{op}^\dagger. 
\end{equation} 


\providecommand{\noopsort}[1]{}\providecommand{\singleletter}[1]{#1}%


\end{document}